\documentstyle[prb,aps,epsf,graphicx]{revtex}

\begin{document}
\draft


\title
{Solution of a class of one-dimensional reaction-diffusion models in disordered media \footnote{To appear in Physical Review B}}

\author{M. Mobilia\footnote{email:mauro.mobilia@epfl.ch} and P.-A. Bares\footnote{email:pierre-antoine.bares@epfl.ch}}
\address
{Institute of Theoretical Physics, Swiss Federal Institute of Technology
  Lausanne, CH-1015 Lausanne  EPFL, Switzerland }

\date{\today} 

\maketitle


\begin{abstract}
We study a one-dimensional class of reaction-diffusion models on a $10-$parameters manifold. The equations of motion of the correlation functions close on this manifold. We compute exactly 
 the long-time behaviour of the density and correlation functions for  {\it quenched} disordered systems. The {\it quenched} disorder consists of disconnected domains of reaction.
We first consider the case where the disorder comprizes  a superposition, with different probabilistic weights, of finite segments, with {\it periodic boundary conditions}. We then pass to the case of finite segments with {\it open boundary conditions}: we solve the ordered dynamics on a open lattice with help of the Dynamical Matrix Ansatz (DMA) and investigate further its  disordered version.

\end{abstract}
\section{Introduction}

In this work we investigate in detail some exact properties, in ordered and disordered media, of a wide class of one-dimensional single-species reaction-diffusion models, which play a central role in the description of interacting many-particle systems in physics, chemistry, biology, etc \cite{Privman} . We consider systems on the  $10-parameter$ manifold on which ({\it in arbitrary dimensions}) the equations of motion of correlation functions close, and the dynamics is soluble.

So far, most theoretical studies on Reaction-Diffusion models consider infinite, ordered media (see e.g \cite{Schutzrev,Grynberg1,Alcaraz} and references therein). We show that for times that exceed the typical diffusion time needed to span the size of the sample, i.e. in the regime where $L^{2}/t \ll 1$, (where $L$ denotes the size of the lattice chain and $t$ the time, see below), finite-size effects dramatically affect the dynamics. A {\it (quenched-)disordered} system consisting of  a random collection of reaction-domains (each segment, of length ${\cal L}$ weighted  with a probablity $\omega ({\cal L})$) of varying sizes, when $\omega ({\cal L})$ is a decreasing exponential, the density exhibits a stretched exponential decay in the diffusion-limited pair annihilation and diffusion-limited coagulation models \cite{Mandache}. This exponential decay  also applies to  the autocorrelation function of symmetric exclusion process (SEP) \cite{Grynberg}. The  picture is realistic for  media, where disorder fractures the medium into disconnected reaction zones. The stretched exponential decay of the density of the photogenerated excitons in $(MX)$ and $(TMMC)$ chains is an experimentally established phenomenon \cite{Kuroda}.
For other distributions $\omega({\cal L})$ and other reaction-diffusion systems the situation is more involved. However, in the regime where $L^{2}/t\ll 1$, the {\it quenched}-disorder always affects the dynamics. Below, we study the density and the two-point non-instantaneous  correlation functions for a class of reaction-diffusion systems in presence of {\it quenched disorder}.

This paper is organized as follows: in the first part we study reaction-diffusion processes occuring on the {\it soluble} $10-$parametric manifold on which the equations of motion of the correlation functions are closed. 
In section II  we determine this manifold as well as the equations of motions of the correlation functions. In section III.A  we obtain the density and non-instantaneous two-point correlation functions for a finite and periodic chain.
In section III.B, via the {\it Dynamical Matrix Ansatz} (DMA), we generalize previous works by Stinchcombe and Sch\"utz \cite{DMA}, on the soluble
 $10-$parametric manifold and obtain original results. The approach developped allows the exact computation of correlation functions on an open chain of arbitrary length, with injection and evacuation of particles at the boundaries. In the thermodynamic limit we compute the exact density and the non-instantaneous correlation functions. As a by-product we generalize the ``{\it Bethe Ansatz}'' equations previously obtained by Stinchcombe and Sch\"utz for the isotropic chain with symmetry-breaking boundary fields. 
In section IV,  we introduce the disordered systems which we consider in the sequel.
In section V, we take advantage of the results of section III.A to study the dynamics of quenched-disordered systems, where the quenched-disorder consist in a  superposition of {\it periodic} segments with probabilistic weight $\omega({\cal L})$. We consider the cases where  $\omega({\cal L})$ is exponential  and where it is an algebraic function.
In section VI  we use the original  results obtained in  section III.B to study the dynamics of quenched-disordered systems, where the quenched-disorder consist in a probabilistic superposition of {\it open} segments with probabilistic weight $\omega({\cal L})$. In section VII, we illustrate previous discussions by considering two physically motivated  examples.

\section{The Formalism}
It is known that models of stochastic hard-core particles are soluble on some manifold on wich the equations of motion of their correlation functions close \cite{Schutz}.

Consider an hypercubic lattice of dimension $d$ with $N$ sites ($N=L^d$), where 
$L$ represents the linear dimension of the hypercube. 
On the lattice, local bimolecular reactions between single-species particle  $A$, with hard-core, take place. Each site can be empty (denoted by the symbol $0$)  
or occupied at most by a particle of type $A$
denoted in the following by the
index $1$ .
The reactions are specified by the transition rates
$\Gamma_{\alpha \beta}^{\gamma \delta}$, where $\alpha,\beta,\gamma,
\delta=0,1$: $\forall (\alpha, \beta)\neq (\gamma, \delta),  \Gamma_{\alpha \beta}^{\gamma \delta}: \alpha + \beta \longrightarrow \gamma +\delta$.

Probability conservation implies $\Gamma_{\alpha \beta}^{\alpha \beta}=-\sum_{(\alpha,\beta)\neq(\alpha',\beta')}
\Gamma_{\alpha \beta}^{\alpha' \beta'}$and
$\Gamma_{\alpha \beta}^{\gamma \delta}\geq 0, \; \forall (\alpha, \beta)\neq (\gamma, \delta)
$
The state of the system is represented by the ket $|P(t)\rangle=\sum_{\{n\}}P(\{n\},t)|\{n\} \rangle$,
where the sum runs over the $2^N, N=L^d$ configurations.
At site $i$ the local state is specified by the ket 
$|n_i\rangle =(1 \; 0 )^{T}$ if the site 
$i$ is empty,  $|n_i\rangle =(0 \; 1 )^{T}$ if the site $i$ is occupied by 
a particle of type $A$ 
($1$) .

We define the {\it left vacuum}  $\langle \widetilde \chi|\equiv \sum_{\{n\}} \langle\{n\} |$, which local representation is $\langle \widetilde{\chi}|=(1 \; 1)\otimes(1 \; 1)$.

It is by now well established that a master equation can be rewritten formally as an imaginary time Schr\"odinger equation: $\frac{\partial}{\partial t}|P(t)\rangle = -H |P(t)\rangle,$
where $H$ is the {\it Stochastic Hamiltonian} which governs the dynamics of the system. In general, it is neither hermitian nor normal. Its construction from the master equation is a standard procedure (see e.g. \cite{Privman,Schutzrev}). The evolution operator $H$ acts locally on two adjacent sites, $H=\sum_{m}H_{m,m+1}$ and because of probability conservation, we have $\langle \widetilde{\chi}|H=0$ . The explicit form of the $H$ considered here can be found, e.g., in Ref.\cite{Schutzrev,Schutz}.

Below we shall assume an initial state $|P(0)\rangle$ and investigate the
expectation value of an operator $O$:$\langle O \rangle(t)\equiv \langle \widetilde \chi|O e^{-Ht}|P(0)\rangle$.
For general single-species bimolecular reaction-diffusion systems, there are $12$ independent parameters \cite{Schutz}. If one imposes to these parameters the $2$ following constraints:

\begin{eqnarray}
\label{eq.0.7}
 \Gamma_{0 0}^{1 0}+ \Gamma_{0 0}^{1 1}-\left(\Gamma_{1 1}^{0 0}+\Gamma_{1 1}^{0 1}  \right)= \Gamma_{0 1}^{1 0} + \Gamma_{0 1}^{1 1}-\left(\Gamma_{1 0}^{0 0}+\Gamma_{1 0}^{0 1}  \right) \;\;;\;\;
 \Gamma_{0 0}^{0 1}+ \Gamma_{0 0}^{1 1}-\left(\Gamma_{1 1}^{0 0}+\Gamma_{1 1}^{1 0}  \right)= \Gamma_{1 0}^{0 1} + \Gamma_{1 0}^{1 1}-\left(\Gamma_{0 1}^{0 0}+\Gamma_{0 1}^{1 0}  \right),
\end{eqnarray}
the equations of motion of the correlation functions close and the system is formally soluble (in {\it arbitrary} dimensions). It is useful to point out that these solvability constraints allow, through a similarity transformation\cite{Alcaraz,Schutz}, to transform the equivalent quantum chain , into another quantum chain sharing the same {\it eigenspectra} of an $XXZ$ quantum chain with surface fields, of the type solved in Ref.\cite{Alacaraz2}.
\section{Density and non-instantaneous correlation functions for ordered case, in one-spatial
dimension on a finite lattice}
For the sequel, in order to compute the sampling average of quantities such as the density and the
non-instantaneous correlation functions we need to know the expression of these quantities on a
finite lattice. In the following subsection we briefly derive the latter expressions on a periodic
finite lattice. In the subsection III.B we compute, as original results, the same expressions on an open lattice with
help of {\it Dynamical Matrix Ansatz}. 
\subsection{The periodic case}
For the class of systems which are described on a $10$-parameter manifold\footnote{This soluble $10-$parametric manifold includes stochastic systems such as Glauber and voter models, the Symmetric Exclusion Process (SEP). }, according to the {\it solubility constraints} (\ref{eq.0.7}), the equation of motion of the density, for a periodic chain of ${\cal L}\leq L$ sites, in one-spatial dimension, reads \cite{Schutz}:
\begin{eqnarray}
\label{eq.0.8}
\frac{d}{dt}\langle n_{m} \rangle (t) = \frac{d}{dt}\langle \widetilde\chi|n_{m}e^{-Ht}|P(0)\rangle = A+B\langle n_{m} \rangle (t)+C\langle n_{m+1} \rangle (t) + D\langle n_{m-1} \rangle (t)
\end{eqnarray}
where we have defined:
\begin{eqnarray}
\label{eq.0.9}
A&\equiv& 2\Gamma_{0 0}^{1 1}+\Gamma_{0 0}^{1 0} + \Gamma_{0 0}^{0 1};\;\;
B\equiv -\left(2\Gamma_{0 0}^{1 1} + \Gamma_{0 0}^{1 0}+\Gamma_{0 0}^{0 1}+ \Gamma_{1 0}^{0 0} + \Gamma_{0 1}^{0 0} +  \Gamma_{1 0}^{0 1} +  \Gamma_{0 1}^{1 0}\right) \nonumber\\
C&\equiv& \Gamma_{0 1}^{1 0} +  \Gamma^{1 1}_{0 1}-\left(\Gamma_{0 0}^{1 1} + \Gamma_{0 0}^{1 0} \right);\;\;
D\equiv \Gamma_{1 0}^{0 1} +  \Gamma^{1 1}_{1 0}-\left(\Gamma_{0 0}^{1 1} + \Gamma_{0 0}^{0 1} \right) ;\;\; 
E\equiv \pm\sqrt{ C D} ;\;\;
\mu \equiv \pm\sqrt{\frac{D}{C}}
\end{eqnarray}

In this work we focus on the case where $E>0$ (which also implies  $\mu>0$).

Solving the  equations of motion obtained  from (\ref{eq.0.7}-\ref{eq.0.9}), we obtain in the ordered case (for integer $m$ and $m'$ and for {\it periodic segments of length ${\cal L}\leq L$}): \\
$\langle n_{m}(t)\rangle_{{\cal L}}=
\frac{2e^{B t}}{{\cal L}}\sum_{m'}\sum_{ p\in 1.B.Z.} \left(\langle n_{m'}(0)\rangle_{{\cal L}} -\langle n_{m'}(\infty)\rangle_{{\cal L}} \right)\mu^{m-m'}(sign E)^{m-m'}\cos \left[p(m-m')\right]e^{2|E|t \cos p}, $

where $\langle n_{m'}(\infty)\rangle\equiv \phi (1-\delta_{A,0})(1-\delta_{B+C+D,0})$, with $\phi\equiv -\frac{A}{B+C+D}$, and  the sum $m'$ runs over all the {\it integers} belonging to the segments
 of length ${\cal L}$ under consideration. The sum over  $p$ runs over the first Brillouin zone ($1.B.Z.$): $p=\frac{2n\pi}{{\cal L}},\;\;n=0,\dots,{\cal L}-1$.

With the same notations, we have for the non-instantaneous two-point correlation functions: $\langle n_{m}(t) n_{l}(0)\rangle_{{\cal L}}=
\frac{2e^{B t}}{{\cal L}}\sum_{m'}\sum_{ p\in 1.B.Z.} \left(\langle n_{m'}(0) n_{l}(0)\rangle_{{\cal L}}-\langle n_{m'}(\infty) n_{l}(0)\rangle_{{\cal L}} \right)\mu^{m-m'}(sign E)^{m-m'}\cos \left[p(m-m')\right]e^{2|E|t \cos p}$, where $\langle n_{m'}(\infty) n_{l}(0)\rangle=\langle n_{m'}(\infty)\rangle$.

It is useful for the sequel to obtain the long-time behaviour of these quantities in the 
regime where ${\cal L}^{2}/|E|t\sim u=L^{2}/|E|t\sim1$, with ${\cal L}\approx L \gg 1 $ 
and $|E|t\gg 1$, for the density we have :
\begin{eqnarray}
\label{eq.0.15.0}
\langle n_{m}\rangle(t)-\langle n_{m}\rangle(\infty)
=\sum_{m'}\mu^{m-m'}\left(\langle n_{m'} (0)\rangle_{{\cal L}} -\langle n_{m'} (\infty)\rangle_{{\cal L}} \right)\frac{e^{(B+2|E|)t-\frac{(m-m')^{2}}{4|E|t}}}{2\sqrt{\pi|E|t}}
\end{eqnarray}
The long-time behaviour (${\cal L}^{2}/|E|t\sim u=L^{2}/|E|t\sim1$, with ${\cal L}\approx L \gg 1 
$ and $|E|t\gg 1$) of the  two-point correlation functions reads
\begin{eqnarray}
\label{eq.0.16.0}
\langle n_{x}(t)n_{x_{0}}(0)\rangle
=\sum_{y}\frac{\mu^{x-y} e^{(B+2|E|)t-\frac{(x-y)^{2}}{4|E|t}}}{2\sqrt{\pi|E|t}}\left(\langle n_{y}(0)n_{x_{0}}(0)\rangle_{{\cal L}} -
\langle n_{y}(\infty)n_{x_{0}}(0)\rangle_{{\cal L}} \right)
\end{eqnarray}
\subsection{Density and non-instantaneous correlation function for the ordered case with open boundary conditions}

It is instructive to consider the physically motivated case of a disordered lattice with open boundary conditions. To investigate this situation, we begin with an ordered lattice of length $L$ and open boundary conditions: we assume that at site $1$  particles can be injected in the system with rate $\widetilde{\alpha}$ and evacuated with rate $\widetilde{\gamma}$. At site $L$  particles can be injected in the system with rate $\widetilde{\delta}$ and evacuated with rate $\widetilde{\beta}$.
We apply  the {\it Dynamical Matrix Ansatz} (DMA) introduced by Stinchcombe and Sch\"utz to study the ordered {\it Symmetric Exclusion Process} (SEP) with open boundary conditions \cite{DMA}. 

For two-states models, the {\it Dynamical Matrix Ansatz} assumes that 
 the probability can be encoded by \cite{DMA} $|P(t)\rangle =
 \langle\langle W|\left\{\prod_{j=1}^{L}(\underline{E}(t)+\underline{D}(t)
 \sigma_{j}^{-})|\Uparrow\rangle\right\}|V\rangle\rangle \frac{1}{Z_{L}},$
where $Z_{L}\equiv\langle\langle W|\underline{C}^{L}|V\rangle\rangle$ is the normalization constant, $\underline{C}=\underline{E}(t)+\underline{D}(t)$, and $\sigma_{j}^{-}$ denotes the usual Pauli matrix acting on site $j$. We have introduced the {\it vectors} $\langle\langle W| $ and $|V\rangle\rangle  $ on which act the (infinite and time-dependent) matrices $\underline{E}(t)$ and $\underline{D}(t)$ (not to be confused with the combination of rates (\ref{eq.0.9})). To investigate  the dynamics, we associate a {\it spin-down} to a particle and a {\it spin-up} to a vacancy. In so doing, the ferromagnetic ground-state $|\Uparrow\rangle$ corresponds to an empty lattice.

In this formulation, correlation functions read:

 $\langle n_{j_{1}}(t) n_{j_{2}}(t)\dots  n_{j_{m}}(t) \rangle=
\langle\langle W|\underline{C}^{j_{1}-1} \underline{D}(t) \underline{C}^{j_{2}-j_{1}-1}
 \underline{D}(t) \dots \underline{C}^{L-j_{m}}|V\rangle\rangle \frac{1}{Z_{L}}$, 
 $ \;\; j_{m}>j_{m-1}>\dots >j_{1}$.

The master equation, in its {\it Hamiltonian} formulation, reads
$\left(\frac{d}{dt}+H\right)|P(t)\rangle = 0$, where $H=\sum_{j=1}^{L-1}H_{j,j+1}+b_{1}+b_{L}$.
The first term in this expression corresponds to the {\it bulk} contribution, while the terms $b_{1}$ and $b_{L}$ correspond to the boundary at site $1$ (injection with rate $\widetilde{\alpha}$ and evacuation with rate $\widetilde{\gamma}$ ) and $L$ (injection with rate $\widetilde{\beta}$ and evacuation with rate $\widetilde{\delta}$ ), respectively.

Proceding as in Ref.\cite{Schutzrev,DMA,Schutz1}, on the $10-$parametric manifold (\ref{eq.0.7}), we obtain the bulk contribution to the master equation: 
\begin{eqnarray}
\label{eq.DMA.8}
\frac{d \underline{D}(t)}{dt}&=&(\sigma_{1}+\sigma_{4})\underline{C}-(\sigma_{3}+\sigma_{5})\underline{D}(t)
+\sigma_{2}\underline{C} \underline{D}(t) \underline{C}^{-1}+\sigma_{6}\underline{C}^{-1} \underline{D}(t) \underline{C} 
\end{eqnarray}
\begin{eqnarray}
\label{eq.DMA.8.0}
\underline{S}(t)&=&(\sigma_{1}-\sigma_{4})\underline{C}+(\sigma_{5}-\sigma_{3})\underline{D}(t) +\sigma_{2}\underline{C}\underline{D}(t)\underline{C}^{-1}
-\sigma_{6}\underline{C}^{-1}\underline{D}(t)\underline{C}
\end{eqnarray}
\begin{eqnarray}
\label{eq.DMA.8.1}
\sigma_{7}\underline{D}(t)^{2}(t)&=&\sigma_{8}\underline{C}\underline{D}(t)+\sigma_{9}\underline{D}(t)\underline{C}-\Gamma_{0 0}^{1 1}\underline{C}^{2}+\sigma_{2}\underline{C}\underline{D}(t)\underline{C}^{-1}\underline{D}(t)+\sigma_{6}\underline{D}(t)\underline{C}^{-1}\underline{D}(t)\underline{C},
\end{eqnarray}
Similarly, the boundary terms give rise to the following equations:
\begin{eqnarray*}
\label{eq.DMA.8.2}
0=\langle\langle W|\left\{(2\sigma_{4}-\widetilde{\alpha})\underline{C}-(2\sigma_{5}-\widetilde{\alpha}-\widetilde{\gamma})\underline{D}(t)+2\sigma_{6}\underline{C}\underline{D}(t)\underline{C}^{-1}\right\}
=\left\{(2\sigma_{1}-\widetilde{\delta})\underline{C}-(2\sigma_{3}-\widetilde{\beta}-\widetilde{\delta})\underline{D}(t)+2\sigma_{2}\underline{C}\underline{D}(t)\underline{C}^{-1}\right\}|V\rangle\rangle,
\end{eqnarray*}
where
\begin{eqnarray}
\label{eq.DMA.9}
&&\sigma_{1}\equiv \Gamma_{0 0}^{1 1}+\Gamma_{0 0}^{1 0}\geq 0\;\;;\;\;
\sigma_{2}\equiv \Gamma_{0 1}^{1 0}+\Gamma_{0 1}^{1 1}-\sigma_{1}\equiv C\;\;;\;\;
\sigma_{3}\equiv \Gamma_{1 0}^{0 0}+\Gamma_{1 0}^{0 1}+ \sigma_{1}\geq 0\nonumber\\
\sigma_{4}&\equiv& \Gamma_{0 0}^{1 1}+\Gamma_{0 0}^{0 1}\geq 0\;\;;\;\;
\sigma_{5}\equiv \Gamma_{0 1}^{0 0}+\Gamma_{0 1}^{1 0}+ \sigma_{4}\geq 0\;\;;\;\;
\sigma_{6}\equiv \Gamma_{1 0}^{0 1}+\Gamma_{1 0}^{1 1}-\sigma_{4}\equiv D \nonumber\\
\sigma_{7}&\equiv& \sigma_{3}+\sigma_{5}-(\Gamma_{1 1}^{0 0}+ \Gamma_{1 1}^{0 1}+\Gamma_{1 1}^{1
0} + \Gamma_{1 0}^{ 1 1}+\Gamma_{0 1}^{ 1 1}+ \Gamma_{0 0}^{ 1 1} ) \;\;;\;\;
\sigma_{8}\equiv \sigma_{1}+\Gamma_{0 0}^{1 1}- \Gamma_{0 1}^{1 1}\;\;;\;\;
\sigma_{9}\equiv \sigma_{4}+\Gamma_{0 0}^{1 1}- \Gamma_{1 0}^{1 1},
\end{eqnarray}
with $\sigma_{2}=C,\sigma_{6}=D, \sigma_{1}+\sigma_{4}=A$ and $\sigma_{3}+\sigma_{5}=-B$, where 
$A,B,C,D$ are the quantities defined in (\ref{eq.0.9}).
For the SEP model, considered in \cite{Schutzrev,DMA,Sosomato}, we have
 $\Gamma_{0 1}^{1 0}=\Gamma_{1 0}^{0 1}=\Gamma>0$ and thus with $\sigma_{2}=\sigma_{3}=
 \sigma_{5}=\sigma_{6}=\frac{\sigma_{7}}{2}=\Gamma$ and $\sigma_{1}=\sigma_{4}=
 \sigma_{8}=\sigma_{9}=0$, we recover previous results Ref.\cite{Schutzrev,DMA,Sosomato}.

To study the dynamical equations (\ref{eq.DMA.8}) with boundary conditions 
(\ref{eq.DMA.9}), it is useful to introduce the following notation: 
$\underline{D}_{j}(t)\equiv \underline{C}^{j-1}\underline{D}(t)\underline{C}^{-j}$, and so the correlation functions read $\langle n_{j_{1}}(t) n_{j_{2}}(t)\dots  n_{j_{m}}(t) \rangle=
\langle\langle W|\underline{D}_{j_{1}}(t)\underline{D}_{j_{2}}(t)
 \dots \underline{D}_{j_{m}} (t) \underline{C}^{L}|V\rangle\rangle \frac{1}{Z_{L}}$, $\;\; 
 j_{m}>j_{m-1}>\dots >j_{1}$.
For the study of the dynamics of the system we can then introduce the Fourier  transform 
of $\underline{D}(t)_{j}(t)$ (for $p\neq 0$): ${\cal D}_{p}(t)\equiv \sum_{j}e^{ipj}\underline{D}_{j}(t)$.
 Seeking a solution of the dynamical equation (\ref{eq.DMA.8}) (for $p\neq 0$) 
of the form ${\cal D}_{p}(t)=e^{-\epsilon_{p}t}{\cal D}_{p}(t=0) $, we 
find $\epsilon_{p}=(\sigma_{3}+\sigma_{5})-\sigma_{2}e^{-ip}-\sigma_{6}e^{ip}$. The equation (\ref{eq.DMA.8.1}) implies the relation: ${\cal D}_{p_{1}}(t){\cal D}_{p_{2}}(t)=
{\cal S}_{p_{2},p_{1}}{\cal D}_{p_{2}}(t){\cal D}_{p_{1}}(t)$,
where  ${\cal S}_{p_{2},p_{1}}$ is the amplitude of the {\it scattering matrix} which plays a central role in the theory of integrable models.
Here, we have
\begin{eqnarray}
\label{eq.DMA.17}
{\cal S}_{p_{2},p_{1}}=-\left(\frac{\sigma_{6}+ \sigma_{2}e^{i(p_{1}+p_{2})} -\sigma_{7}e^{ip_{2}}}{\sigma_{6}+ \sigma_{2}e^{i(p_{1}+p_{2})} -\sigma_{7}e^{ip_{1}} }\right), \;\; (p_{1},p_{2})\neq(0,0)
\end{eqnarray}
For the SEP model we recover \cite{Schutzrev,DMA} the {\it scattering amplitude of the isotropic Heisenberg chain}

With the definitions,
\begin{eqnarray}
\label{eq.DMA.19}
2\alpha \equiv\frac{\widetilde{\alpha}-2\sigma_{4}}{2\sigma_{6}} \;;\;
2\beta \equiv 1+\frac{\widetilde{\beta}+2(\sigma_{1}-\sigma_{3})}{2\sigma_{2}} \;;\;
2\gamma \equiv 1+ \frac{\widetilde{\gamma}+2(\sigma_{4}-\sigma_{5})}{2\sigma_{6}} \;;\;
2\delta \equiv \frac{\widetilde{\delta} -2\sigma_{1}}{2\sigma_{2}},
\end{eqnarray}
it follows from the boundary conditions that 
\begin{eqnarray}
\label{eq.DMA.20}
{\cal T}(p)\equiv\langle\langle W| {\cal D}_{p}\underline{C}^{L}|V\rangle\rangle=-\frac{f_{p}(\alpha,\gamma)}{f_{-p}(\alpha,\gamma)}e^{2ip}{\cal T}(-p)=
-\frac{f_{p}(\beta,\delta)}{f_{-p}(\beta,\delta)}e^{2ipL}{\cal T}(-p)
, (p\neq 0),
\end{eqnarray}
where $f_{p}(a,b)\equiv 2(a+b)-1+e^{ip}$.
These equations have as solution: ${\cal T}(p)=\sum_{l>0}a_{l}\left(e^{ipl}-e^{ip(2-l)}\frac{f_{-p}(\alpha,\gamma)}{f_{p}(\alpha,\gamma)}\right)=
\sum_{l>0}a_{l}\left(e^{ipl}-e^{ip(2L-l)}\frac{f_{p}(\beta,\delta)}{f_{-p}(\beta,\delta)}\right)$.
Consistency implies the following relation which quantizes the momenta $p$'s, as in  quantum Many-Body problems:
\begin{eqnarray}
\label{eq.DMA.22}
B(p)\equiv \frac{f_{-p}(\alpha,\gamma)f_{-p}(\beta,\delta)}
{f_{p}(\alpha,\gamma)f_{p}(\beta,\delta)}= e^{2ip(L-1)}
\end{eqnarray}
and more generally, for the $m-$points correlation functions, we have $ e^{2ip_{j}(L-1)}=B_{p_{j}}\prod_{k=1 (k \neq j)}^{m}
\frac{{\cal S}(p_{j},p_{k})}{{\cal S}(p_{j},-p_{k})}$.
Equations (\ref{eq.DMA.22}) and its $m-$points generalization are the analog of the {\it Bethe Ansatz} equations for quantum Many-Body problems \cite{Schutzrev,DMA}.

Correlation functions are then obtained as follows: 

$\langle n_{j_{1}}(t)\dots  n_{j_{m}}(t)\rangle -
\langle n_{j_{1}}(\infty)\dots  n_{j_{m}}(\infty)\rangle=\int\dots\int
\prod_{i=1}^{m}\left[dp_{i} e^{-ip_{i}-\epsilon_{p_{i}}t}\right] {\cal T}(p_{1},\dots,p_{m})$,
 $\nonumber\\ j_{m}>\dots>j_{1}\;;\; p_{i}\neq 0\;;\; 1\leq i\leq m$.
Where ${\cal T}(p_{1},\dots,p_{m})\equiv \langle\langle W| {\cal D}_{p_{1}}(0) \dots
 {\cal D}_{p_{m}}(0)|V\rangle\rangle\frac{1}{Z_{L}}$ encodes the initial state of the system.

For finite systems the {\it integration} over the momenta,  should be carried over $p's$ satisfying the relations (\ref{eq.DMA.22}). For infinite system (when $L\rightarrow \infty$) these relations (\ref{eq.DMA.22}) can always be fulfilled (there is a dense set of solutions of these equations) and the integration over the $p$'s,  runs over the first Brillouin zone (with exclusion of $p=0$ which only contributes to the static part of the correlation functions).

We will now focus on the density, which reads (for large $L$):
\begin{eqnarray}
\label{eq.DMA.25}
\langle n_{x}(t)\rangle-\langle n_{x}(\infty)\rangle=
\sum_{l>0}\left(\langle n_{l}(\infty)\rangle - \langle n_{l}(0)\rangle\right)\int \frac{dp}{2\pi} e^{-ipx-\epsilon_{p}t}\left(e^{i{p}l} -
e^{ip(2-l)}\frac{f_{-p}(\alpha,\gamma)}{f_{p}(\alpha,\gamma)}\right), (p\neq 0)
\end{eqnarray}
With the boundary conditions, we obtain for the stationary density: 

i) For the SEP model, where $\sigma_{3}+\sigma_{5}=\sigma_{2}+\sigma_{6}$ (i.e. $|B|=C+D$) 
with $\sigma_{2}=\sigma_{6}>0$ and $\sigma_{1}+\sigma_{4}=A=0$, we recover the results of 
Stinchcombe and Sch\"utz \cite{Schutzrev,DMA}.

ii) For the other cases on the $10-$parametric manifold (\ref{eq.0.7}), for which $|B|\neq C+D$, i.e. $\sigma_{3}+\sigma_{5}\neq \sigma_{2}+\sigma_{6}$,   and  thus $A=\sigma_{1}+\sigma_{4}\neq 0$ :
\begin{eqnarray}
\label{eq.DMA.26}
\langle n_{1}(\infty)\rangle&=&
\phi+\frac{-\sigma_{2}\tau_{4}\eta+\tau_{2}(\tau_{3}-\sigma_{6}\eta^{L-2})}
{(\eta\sigma_{2}+\tau_{1})(\tau_{3}+\sigma_{6}\eta^{L-2})-\sigma_{2}\sigma_{6}\eta^{L-1}}\;\;;\;\;
\langle n_{L}(\infty)\rangle=\phi+\frac{\tau_{4}-\sigma_{6}\eta^{L-2}(\langle n_{1}(\infty)\rangle  -\phi)}{\tau_{3}+\sigma_{6}\eta^{L-2}}
\nonumber\\
\langle n_{x}(\infty)\rangle&=& (\langle n_{1}(\infty)\rangle+\langle n_{L}(\infty)\rangle-2\phi)\eta^{x-1}+\phi, \;\; 1<x<L,
\end{eqnarray}
where, $\phi\equiv\frac{\sigma_{1}+\sigma_{4}}{\sigma_{3}+\sigma_{5}-\sigma_{2}-\sigma_{6} }\;;\;$
$\eta\equiv\frac{\sigma_{3}+\sigma_{5}-\sqrt{(\sigma_{3}+\sigma_{5})^{2} -4\sigma_{2}\sigma_{6}}}{2\sigma_{2}}\;;\;$
$\tau_{1}\equiv (1-2(\alpha+\gamma))\sigma_{6}-(\sigma_{3}+\sigma_{5});\;\;$
$\tau_{2}\equiv 2\sigma_{4}-\widetilde{\alpha}+\phi\left[2(\sigma_{6}-\sigma_{5})
+\widetilde{\alpha}+\widetilde{\gamma}\right];\;\;$
$\tau_{3}\equiv (1-2(\beta+\delta))\sigma_{2}-(\sigma_{3}+\sigma_{5});\;\;$ and 
$\tau_{4}\equiv  2\sigma_{1}-\widetilde{\delta}+
\phi\left[2(\sigma_{2}-\sigma_{3})+\widetilde{\beta}+\widetilde{\delta}\right]$

The dynamical part of the density reads, according to (\ref{eq.DMA.25}),

i) For the SEP model,  where $B+C+D=\sigma_{3}+\sigma_{5}-\sigma_{2}-\sigma_{6}=0$ and
$A=\sigma_{1}+\sigma_{4}=0$,  we recover results obtained in references Ref.\cite{Schutzrev,DMA}.

ii)
For the other processes on the $10-$parameter manifold (\ref{eq.0.7}) (for which $\sigma_{3}+\sigma_{5}\neq \sigma_{2}+\sigma_{6}$ with $A\neq 0$ and $\sigma_{2}\sigma_{6}\neq 0$), we have
\begin{eqnarray}
\label{eq.DMA.28}
\langle  n_{x}(t)\rangle-\langle n_{x}(\infty)\rangle&=&
-e^{Bt}\sum_{y=1}^{L}\left(\frac{D}{C}\right)^{\frac{x-y}{2}}\left(
\langle  n_{y}(0)\rangle-\langle n_{y}(\infty)\rangle
\right)\left(I_{x-y}(2|E|t)+\triangle I_{x+y-1}(2|E|t)\right)\nonumber\\&+&
e^{Bt}(1-\triangle^{2})\sum_{y=1}^{L}\sum_{k=1}^{y-1}\triangle^{y-k-1}\left(\frac{D}{C}\right)^{\frac{x-k}{2}}
\left(
\langle  n_{k}(0)\rangle-\langle n_{k}(\infty)\rangle
\right)I_{x+y-1}(2|E|t)
 ,
\end{eqnarray}
where $\triangle\equiv \sqrt{\frac{\sigma_{6}}{\sigma_{2}}}
(1-2(\alpha+\gamma))= \frac{2\sigma_{5}-(\widetilde{\alpha}+\widetilde{\gamma})}{2\sqrt{\sigma_{2}\sigma_{6}}}$.

Similarly, we obtain the expression of the non-instantaneous two-point correlation functions $\langle n_{x}(t)n_{x_{0}}(0)\rangle$ for an initially translationally invariant system with initial density $\langle n_{m}(t=0)\rangle=\rho(0), m=1,\dots,L$. In so doing, we have to replace in (\ref{eq.DMA.28}) $\langle n_{j}(0)\rangle$ with $\rho^{2}(0)(1-\delta_{x,x_{0}})+\rho(0)\delta_{x,x_{0}}$.
\section{The quenched disordered situation}
In the subsequent sections we will exploit the results of the previous section to compute the density and
non-instantaneous correlation functions of systems obeying (\ref{eq.0.7}) and in presence of
 quenched disordered.
More specifically, we assume that the reaction-rates defined in (\ref{eq.0.9}) obey a specific probability distribution.
We assume a distribution, which has already been considered in the study the Glauber-Ising Model
 and other diluted chains (see e.g. \cite{Privman,Droz,Stinch}), and more recently, of
  the symmetric  one-dimensional single-species diffusion-limited reaction 
  $A+\emptyset \leftrightarrow \emptyset+A$ \cite{Grynberg}. 
  The first distribution under consideration is the so-called  ``peaked distribution'' which will
   describe a {\it diluted chain}: ${\cal P}(A,B,C,D)=q^{\hat{\alpha} {\cal L}^{\hat{\beta}}}\delta_{A,A_{1}}\delta_{B,B_{1}}\delta_{C,C_{1}}\delta_{D,D_{1}} + (1-q^{\hat{\alpha} {\cal L}^{\hat{\beta}}} )
   \delta_{A,0}\delta_{B,0}\delta_{C,0}\delta_{D,0}$,
where $0<q<1$ and $A,B,C,D$ have been defined in (\ref{eq.0.9}).
According to this distribution, the chain is broken into discontinuous segments of (finite) length ${\cal L}$, with ${\cal L}+1$ sites. On these segments the reaction-rates are respectively $A_{1}, B_{1}, C_{1}, D_{1}$. Therefore, each realization has a {\it probabilistic weight} $\omega({\cal L})$:
\begin{eqnarray}
\label{eq.1.11}
\omega({\cal L})=q^{\hat{\alpha} {\cal L}^{\hat{\beta}}}, \hat{\alpha}, \hat{\beta} >0,
\end{eqnarray}
if the segments under consideration have {\it periodic boundary conditions}
and a  {\it probabilistic weight} $\omega({\cal L})$:
\begin{eqnarray}
\label{eq.1.11.0}
\omega({\cal L})=\left\{
\begin{array}{l l}
(1-q^{\hat{\alpha}})^2 q^{\hat{\alpha} {\cal L}^{\hat{\beta}} }  &\mbox{, if $1\leq {\cal L} \leq L -2$ }\\
(1-q^{\hat{\alpha}}) q^{\hat{\alpha} {\cal L}^{\hat{\beta}} }   &\mbox{, if $ {\cal L} =L-1$  }
\end{array}
\right.
\end{eqnarray}
if the segments under consideration are  {\it open segments}.

In  Ref.\cite{Grynberg}, such a distribution (\ref{eq.1.11.0}) has been considered for the symmetric $A+\emptyset\leftrightarrow \emptyset+A$ reaction, for an initially homogeneous and uncorrelated system. The authors used the relationship of their model to the Heisenberg chain and its symmetries to compute explicitly the autocorrelation function  $\langle n_{r}(t)n_{r}(0)\rangle$ .

For {\it periodic segments}, we also consider the {\it algebraic} disorder-distribution \cite{Mandache}:
\begin{eqnarray}
\label{eq.1.12}
\omega({\cal L})=\left\{
\begin{array}{l l}
 0 &\mbox{, if ${\cal L}<b $ }\\
 (\gamma -1)b^{\gamma -1}{\cal L}^{-\gamma} &\mbox{, if ${\cal L}>b$; with $\gamma >2$ }
\end{array}
\right.
\end{eqnarray}

In this section we want to extend Grynberg and Stinchcombe's \cite{Grynberg} results 
and calculate explicitly the density and the non-instantaneous two-point correlation functions on 
the $10-$parameter space manifold  (\ref{eq.0.7}) on which the equations of motion of the 
correlation functions are closed. This is first performed for disordered systems of periodic segments
 of length ${\cal L}$  and then for disordered systems of open segments. 

To obtain the density of the disordered systems, which we denote  $\overline{\langle n_{m}\rangle}(t)$, and of the non-instantaneous two-point correlation functions
 $\overline{\langle n_{m}(t)n_{l}(0)\rangle}$, (where the symbol $\overline{\langle \dots \rangle}(t)$ means the average over the time and the disorder), we need the density $\langle n_{m}(t)\rangle_{{\cal L}}$  and the non-instantaneous two-point correlation functions $\langle n_{m}(t)n_{l}(0)\rangle_{{\cal L}}$  for a finite and ordered system  of length ${\cal L}$.
In this section we consider the disorder-averaging of an observable, say $O$, as:
\begin{eqnarray}
\label{eq.1.13}
\overline{\langle O(t)\rangle}\equiv \frac{\sum_{{\cal L}}\omega({\cal L}){\cal L}\langle O(t) \rangle_{{\cal L}}}{\overline{{\cal L}}},
\end{eqnarray}
where $\overline{{\cal L}}\equiv\sum_{{\cal L}}\omega({\cal L}){\cal L} $, and $\langle O(t)\rangle_{{\cal L}}$ denotes, respectively,  the (time-)mean-value of the segment of length ${\cal L}$  and the (time-)mean-value of an observable $O$ on a lattice of size ${\cal L}$. In (\ref{eq.1.13}), the averaging over the samples (i.e. the {\it quenched-disorder}) is carried out  over segments of length ${\cal L}$. It has to be stressed that this summation must be {\it compatible} with the {\it nature} of the observable $O$. In fact, if one is interested in computing the density of particles at site $m$, namely $\overline{\langle n_{m}(t)\rangle}$, one should  carry the summation over all the segments which include the site $m$. In this case, 
$\overline{\langle n_{m}(t)\rangle} =\frac{\sum_{{\cal L}\geq m}\omega({\cal L}){\cal L}\langle n_{m}(t) \rangle_{{\cal L}}}{\sum_{{\cal L}\geq m} \omega({\cal L}){\cal L}} $ and similarly for the correlation functions.

To study the mean value of an observable $O$, which time-average on a finite segment of length ${\cal L}$ is $\langle O(t)\rangle_{{\cal L}} $, we distinguish two regimes:

To study the quantity $\overline{\langle O(t)\rangle}$ (\ref{eq.1.13}), we will distinguish two long-time (i.e. $|E|t\gg 1$) regimes: the regime where $u\equiv L^{2}/|E|t \sim 1$, i.e. for  typical diffusion-time and the regime where $u\equiv L^{2}/|E|t \ll 1$, where time is much larger than the typical diffusion-time. In the latter the finite-size effects due to the disorder sampling procedure dramatically affect the dynamics.
There is also a third regime,   where  $|E|t\gg 1$ and   $u=L^{2}/|E|t \gg 1$, which is less interesting for our purpose, where  we  have $\overline{\langle O(t) \rangle}\sim \langle O(t) \rangle $ \cite{Grynberg,Mandache}. In the sequel we will focus on the regimes $u\sim 1$ and $u\ll 1$ and omit the case $u\gg 1$. 

We are interested in the long-time behaviour of statistical quantities such as density or correlation 
functions (see section II.A) and (\ref{eq.DMA.28}). It has been established 
(see e.g. Ref.\cite{Schutzrev,Grynberg}) that the long-time behaviour of these quantities are
 governed by the {\it small momenta} $p$, ergo by large segments,
  with $1\ll {\cal L}\lesssim L\rightarrow \infty$.
Here the  observables $O$ considered are the density operator at site $j$, i.e. $O=n_{j}$ and
  the operator $n_{j}n_{k}$. For these operators, as
   discussed, the long-time behaviour on a finite segment of length ${\cal L}$ 
   follows from a small momentum expansion\cite{phase-shift} :
    $\langle O(t)\rangle\sim e^{(B+2|E|)t}\sum_{p}\exp\left(-|E|t p^{2}(1+{\cal O}(1/{\cal L}^{2}))\right)$
    , according to the results of section II.A and (\ref{eq.DMA.28}).
     The long-time behaviour ($|E|t \gg 1$) in the presence of quenched-disorder, in 
     the regimes $u\sim 1$ and $u\ll 1$ follows as \cite{Mandache} :
\begin{eqnarray}
\label{eq.1.14}
\overline{\langle O(t)\rangle}
\sim\left\{
\begin{array}{l l}
 \frac{\sum_{{\cal L}\geq\sqrt{\pi^{2}|E|t}}  {\cal L} \omega({\cal L}) \langle O(t)\rangle_{{\cal L}}}{ \sum_{{\cal L}} {\cal L}\omega({\cal L})}
\approx  \langle O(t) \rangle\frac{\sum_{{\cal L}\geq\sqrt{\pi^{2}|E|t}}  {\cal L} \omega({\cal L}) }{ \sum_{{\cal L}} {\cal L} \omega({\cal L})}
 &\mbox{, if $u\sim 1  $ }\\
 \frac{\sum_{{\cal L}<\sqrt{\pi^{2}|E|t}}  {\cal L} \omega({\cal L}) \langle O(t)\rangle_{{\cal L}}}{ \sum_{{\cal L}}{\cal L} \omega({\cal L})}
 &\mbox{, if $u\ll 1  $ }
\end{array}
\right.
\end{eqnarray}

Let us note that the SEP model with free boundary conditions, 
i.e.  $\widetilde{\alpha}=\widetilde{\beta}=\widetilde{\gamma}=\widetilde{\delta}=0 $, 
 is special in the sense that in this case the quantization constraint
  (\ref{eq.DMA.22}) of the momenta $p$'s is always (for segments of arbitrary length ${\cal L}$,
   and not only for the case ${\cal L}\approx L\rightarrow \infty$) fulfilled with
     $p=\frac{2\pi j}{{\cal L}}, \; j=0,1,\dots,{\cal L}-1$ \cite{DMA}.
\section{quenched disorder for periodic segments}

In this section, we compute the density and the non-instantaneous two-point correlation functions on the {\it soluble  $10$-parameteric manifold} for systems with periodic boundary conditions \footnote{
For segments (with periodic boundary conditions  as well as for open segments) caution is required in order to preserve the parity of the processes when considering the quenched-disordered steady-states of the system. As an example of processes which preserve the parity of the number of particles, such as  $A+\emptyset\leftrightarrow \emptyset +A$ and $A+A\leftrightarrow \emptyset + \emptyset $  or the process 
$A+\emptyset\leftrightarrow \emptyset +A$ and $A+A\rightarrow \emptyset + \emptyset $, one has to take into account that for (finite and infinite) segments with initially {\it odd} number of particles there will always remain one particle on each segment (see Ref.\cite{Mandache}). That is the reason for the steady-state (e.g. $\overline{\langle n_{r}(\infty) \rangle}$) to be considered separatly. In order to avoid the difficulties due to {\it residual particles}, we can  restrict to the {\it even subspace}, i.e. $\frac{1}{2}(\openone +\prod_{m}\sigma_{m}^{z})|P(0)\rangle$. Therefore,  we can, e.g.,  only take into account segments with even number of particles and the disordered steady-states correspond to $\overline{\langle \dots \rangle}(t=\infty)=\frac{1}{\overline{{\cal L}}}\sum_{{\cal L}}\omega({\cal L}){\cal L}\langle \dots \rangle_{{\cal L}}(t=\infty)$
}.

For these periodic systems,  we assume  that {\it quenched-disorder} is characterized by two different distributions: the exponential distribution (\ref{eq.1.11}) and the power-law  distribution (\ref{eq.1.12}).

1) We begin with the first distribution (\ref{eq.1.11}) and introduce the following notations:
$\xi\equiv-\frac{1}{\ln q};\;s\equiv\frac{{\cal L}}{\xi};\; \tau \equiv\frac{2|E|t}{\xi^{2}}\;\;;\;\;
{\cal A}\equiv \hat{\alpha} \xi^{\frac{2(\hat{\beta}-1)}
{\hat{\beta}+2}}\left(\frac{4\pi^{2}}{\hat{\alpha}\hat{\beta}}\right)^{\frac{\hat{\beta}}
{\hat{\beta}+2}}+2\pi^{2}\xi^{2\left(\frac{\hat{\beta}-1}{\hat{\beta}+2}\right)} 
\left(\frac{4\pi^{2}}{\hat{\alpha}\hat{\beta}}\right)^{-\frac{2}{\hat{\beta}+2}}\;\;;\;\;$ and 
${\cal B}\equiv \hat{\alpha}\hat{\beta}(\hat{\beta}-1) \left(\frac{4\pi^{2}}{\hat{\alpha}
\hat{\beta}}\right)^{\frac{\hat{\beta}-2}{\hat{\beta}+2}} +12\pi^{2} 
\left(\frac{4\pi^{2}}{\hat{\alpha}\hat{\beta}}\right)^{-\frac{4}{\hat{\beta}+2}}$

1.i) 
To study the density and the correlation functions in this regime, we first expand the expressions 
obtained in section III.A at small momenta (large ${\cal L}$) and then proceed to the disorder-averaging according to (\ref{eq.1.14}) within  a {\it saddle point expansion}.
 
In the regime where $u\ll 1$, we have for the density of a system with a non-uniform initial density profile :
$\langle n_{m}(t=0)\rangle=\langle n_{m}(t=\infty)\rangle(1+\delta_{m,1})$. With $r\equiv m-1$, we have:
\begin{eqnarray}
\label{eq.1.15}
\overline{\langle n_{m}(t)\rangle -\langle n_{m}(\infty)\rangle}\approx
\frac{\sqrt{2\pi}\langle n_{m}(\infty)\rangle e^{r\epsilon}}{\bar{{\cal L}} \xi^{2\left(\frac{\hat{\beta}-1}{\hat{\beta}+2}\right)}}
\frac{\exp\left(-{\cal A}\tau^{\frac{\hat{\beta}}{\hat{\beta}+2}}+(B+2|E|)\frac{\xi^{2}\tau}{2|E|}\right)}{\sqrt{{\cal B}\tau^{\frac{\hat{\beta} -2}{\hat{\beta} +2}}}}
\end{eqnarray}
For the non-instantaneous correlation functions, we also consider a non uniform initial system: $\langle n_{y}(0)  n_{x_{0}}(0)\rangle=\langle n_{y}(\infty)  n_{x_{0}}(0) \rangle(1+\delta_{y,1})(1-\delta_{y,x_{0}})+ \langle n_{x_{0}}(0)\rangle\delta_{y,x_{0}}$. The expression of $\overline{\langle n_{x}(t)  n_{x_{0}}(0)\rangle -\langle n_{r}(\infty)n_{0}(0)\rangle}$ 
with such initial state is obtained from r.h.s. of (\ref{eq.1.15}) in replacing $\langle n_{m}(\infty)\rangle$ with $\mu^{x-1}\left(\langle n_{x_{0}}(\infty)\rangle+\mu^{1-x_{0}}(\langle n_{x_{0}}(0)\rangle)-(1+\delta_{x_{0},1})
\langle n_{x_{0}}(\infty)\rangle \right)$.

1.ii) In the regime where $u\sim 1$,according to (\ref{eq.1.14}), we have for the density
\begin{eqnarray}
\label{eq.1.17}
\overline{\langle n_{m}(t)\rangle -\langle n_{m}(\infty)\rangle}\approx
\left(\frac{\hat{\alpha}}{\xi}\right)^{\frac{2-\hat{\beta}}{\hat{\beta}}}\left(\pi^{2}|E|t\right)^{1-\frac{\hat{\beta}}{2}}e^{-\frac{\hat{\alpha}}{\xi}(\pi^{2}|E|t)^{\frac{\hat{\beta}}{2}}}\left(\langle n_{m}(t)\rangle -\langle n_{m}(\infty)\rangle \right)
\end{eqnarray}
For the non-instantaneous correlation functions, according to (\ref{eq.1.14}), we replace in the r.h.s. of  (\ref{eq.1.17}) $\left(\langle n_{m}(t)\rangle -\langle n_{m}(\infty)\rangle \right)$ with $\left(\langle n_{r}(t)n_{0}(0)\rangle -\langle n_{r}(\infty)n_{0}(0)\rangle \right)$
Expressions $\langle n_{m}(t)\rangle -\langle n_{m}(\infty)\rangle$
and $\langle n_{r}(t)  n_{0}(0)\rangle -\langle n_{r}(\infty)\rangle$ have been  obtained previously (\ref{eq.0.15.0}) and (\ref{eq.0.16.0})  for the (infinite) ordered periodic chain
\\
We now consider the second (algebraic) distribution (\ref{eq.1.12}) for the disorder average, with $\gamma>2$.

2.i) For a system characterized by a non-uniform initial density:
 $\langle n_{m}(t=0)\rangle=\langle n_{m}(\infty)\rangle(1+\delta_{ m,1})$, with $r\equiv m-1$
\begin{eqnarray}
\label{eq.1.19}
\overline{\langle n_{m}(t)\rangle -\langle n_{m}(\infty)\rangle}\approx
\frac{\langle n_{m}(\infty)\rangle e^{r\epsilon}}{\overline{{\cal L}}\xi}
\pi(\gamma-1)b^{\gamma-1}\Gamma\left(\frac{\gamma-1}{2},1\right)\frac{e^{(B+2|E|)t}}{(4\pi^{2}|E|t)^{(\gamma-1)/2}}
\end{eqnarray}
For systems with non-uniform initial states such that $\langle n_{y}(0)  n_{x_{0}}(0)\rangle=\langle n_{y}(\infty)  n_{x_{0}}(0) \rangle(1+\delta_{y,1})(1-\delta_{y,x_{0}})+ \langle n_{x_{0}}(0)\rangle\delta_{y,x_{0}}$, we have to replace in the r.h.s. of (\ref{eq.1.19}) $\langle n_{m}(\infty)\rangle$ with
$\mu^{x-1}\left(\langle n_{x_{0}}(\infty)\rangle+\mu^{1-x_{0}}(\langle n_{x_{0}}(0)\rangle)-(1+\delta_{x_{0},1})
\langle n_{x_{0}}(\infty)\rangle \right)$.

2.ii) In the regime where $u\sim 1$, we have for the density, according to (\ref{eq.1.14}),
\begin{eqnarray}
\label{eq.1.21}
\overline{\langle n_{m}(t)\rangle -\langle n_{m}(\infty)\rangle}\approx
\left(\frac{\gamma-1}{\gamma-2}\right)b^{\gamma-1}(4\pi^{2}|E|t)^{1-\frac{\gamma}{2}}\left(\langle n_{m}(t)\rangle - \langle n_{m}(\infty)\rangle\right)
\end{eqnarray}
For  the non-instantaneous correlation functions, according to (\ref{eq.1.14}), we obtain: $\overline{\langle n_{r}(t)  n_{0}(0)\rangle -\langle n_{r}(\infty)\rangle}\approx
\left(\frac{\gamma-1}{\gamma-2}\right)b^{\gamma-1}(4\pi^{2}|E|t)^{1-\frac{\gamma}{2}}\left(\langle n_{r}(t)n_{0}(0)\rangle - \langle n_{r}(\infty)n_{0}(0)\rangle\right)$.

In these expressions $\langle n_{m}(t)\rangle -\langle n_{m}(\infty)\rangle$
and $\langle n_{r}(t)  n_{0}(0)\rangle -\langle n_{r}(\infty)\rangle$ have been obtained 
in section III.A.

\section{Quenched-disorder for open segments}

In this section (with the same notations as in sections V and VI), taking advantage of the  results of the section IV, we compute density and correlation
 functions  on the {\it soluble  $10$-parametric  manifold} for systems with open boundary conditions.

We consider the effect of quenched-disorder for open segments.
We assume that the system is initially uncorrelated, with  density $\rho(0)$.
For technical convenience, we assume the {\it exponential} distribution (\ref{eq.1.11.0}).
In what follows, the case of (finite) segments with free-boundary conditions (without injection and evacuation of particles at the sites $1$ and ${\cal L}$) corresponds to the case where $\widetilde{\alpha}=\widetilde{\beta}= \widetilde{\gamma}=\widetilde{\delta}=0$ and thus $\alpha=\frac{\sigma_{4}}{2\sigma_{6}},\beta= \frac{1}{2}+\frac{\sigma_{1}-\sigma_{3}}{2\sigma_{2}},\gamma=\frac{1}{2}+\frac{\sigma_{4}-\sigma_{6}}{2\sigma_{6}},\delta=-\frac{\sigma_{1}}{2\sigma_{2}}$.

We introduce the quantities $\xi'_{i}=\xi'_{i}(\hat{\alpha},\hat{\beta})\;\;, i=1,2,3$, where

\begin{eqnarray}
\label{eq.DMA.31}
\xi_{1}'\equiv\pm \ln\left(\frac{D}{C}\right)\;\;,\;\;
\xi_{2}'\equiv\ln|\triangle|\pm \ln\left(\frac{D}{C}\right)\;\;,\;\;
\xi_{3}'\equiv\ln|\eta|\pm \ln\left(\frac{D}{C}\right)
\end{eqnarray}
where, as previously, we consider $D/C=\sigma_{6}/\sigma_{2}>0$ and {\it assume} $\triangle=\frac{2\sigma_{5}-(\widetilde{\alpha}+\widetilde{\gamma})}{2\sigma_{6}}> 0$, for $\sigma_{5}> 0$ and  $DC=\sigma_{6}\sigma_{2}>0$, without further restrictions, we can always {\it assume} that $2\sigma_{5}\geq \widetilde{\alpha}+\widetilde{\gamma}> 0$. In particular, this holds true for {\it free-boundaries} systems, where $\widetilde{\alpha}=\widetilde{\beta}=\widetilde{\gamma}= \widetilde{\delta}=0$.

As in the previous section, we distinguish two regimes:

i) In the first one, $L^{2}/|E|t\ll 1$, the result is obtained via a saddle-point expansion.

To study the density and the correlation functions in this regime, we first expand the expressions (\ref{eq.DMA.28}) and its  correlation functions for small momenta (large ${\cal L}$) and then proceed to the disorder-averaging according to (\ref{eq.1.14}) within  a saddle point expansion. 
In so doing, the saddle points  $s_{0}=s_{0}(\xi'_{i})\;\;,\;\;i=1,2,3$ are  solutions of the {\it saddle-point equation}: $\xi\xi'_{i}-\hat{\alpha}\hat{\beta}(s_{0}\xi)^{\hat{\beta}-1}+\frac{4\pi^{2}\tau}{s_{0}^{3}}=0\;\;,\;\;i=1,2,3$
For the case $\hat{\beta}=1$, we have
$s_{0}(\xi_{i}')=\left(\frac{4\pi^{2}\tau}{\hat{\alpha}-\xi\xi_{i}'}\right)^{1/3}$. 

The long-time behaviour of the density after averaging over the time and the disorder reads (for $x\ll L$):
\begin{eqnarray}
\label{eq.DMA.33}
\overline{\langle n_{x}(t)\rangle - \langle n_{x}(\infty)\rangle}\approx
 \tau^{\nu(\xi_{i}')}\exp\left[\frac{(B+2|E|)\xi^{2} \tau}{2|E|}+max_{\left\{\xi_{i}'\right\}_{i=1, 2, 3}}\left\{\xi\xi_{i}' s_{0}(\xi'_{i}) - \hat{\alpha}\xi^{\hat{\beta}-1}s_{0}(\xi'_{i})^{\hat{\beta}}-\frac{2\pi^{2}\tau}{s_{0}^{2}(\xi_{i}')}\right\}  \right],
\end{eqnarray}
where $\nu=\nu(\xi_{i}')$ is an exponent which depends on the parameters $\xi_{i}'$ . On the  r.h.s. of (\ref{eq.DMA.33}), in the term $max_{\left\{\xi_{i}'\right\}_{i=1, 2, 3}}\left\{\dots\right\}$, we {\it select} the maxima (over the $\xi_{i}'$) of the function $f(s_{0}(\xi_{i}'))= \xi\xi'_{i}s_{0}(\xi_{i}')-\hat{\alpha}\xi^{\hat{\beta}-1}s_{0}(\xi'_{i})^{\hat{\beta}}-\frac{2\pi^{2}\tau}{s_{0}^{2}(\xi_{i}')},\;\;(i=1,2,3) $, where the $s_{0}(\xi_{i}')$ are the solutions of the saddle-point equation.

We see from the formula (\ref{eq.DMA.33}) that the effect of the {\it open boundary} appears through the term $\xi'_{i}$ (\ref{eq.DMA.31}) and can affect the dynamics with respect to the case of segments with periodic boundary conditions. 

For the case where $\hat{\beta}=1$  all the computations can be explicitly carried out as follows.

We define $\xi'$ according to $\left\{\hat{\alpha}-\xi\xi'\right\}^{2/3}\equiv
max_{\{\xi_{i}'\}_{i=1,2,3}}\left\{\hat{\alpha}-\xi\xi_{i}'\right\}^{2/3}$. To compute the density, we then distinguish two cases (with $\hat{\alpha}\neq \xi\xi'$). First, if $\xi'\neq 0$ (or $\xi'= 0$ and $\triangle=1$, as for the SEP model).
\begin{eqnarray}
\label{eq.DMA.34}
\overline{\langle n_{x}(t)\rangle - \langle n_{x}(\infty)\rangle}\approx
\frac{1}{\overline{{\cal L}}\xi}\sqrt{\frac{2\pi}{3}}\left(\frac{4\pi^{2}\tau}{(\hat{\alpha}-\xi\xi')^{4}}\right)^{1/6}\exp\left(-\frac{3(\hat{\alpha}-\xi\xi')^{2/3}}{2}(4\pi^{2}\tau)^{1/3}+\left(\frac{B+2|E|}{2|E|}\right) \xi^{2} \tau\right)
\end{eqnarray}

If $\xi'=0$, $\triangle \neq 1$  and $\hat{\beta}=1$, for the situation considered here (systems initially uncorrelated with initial density $\rho(0)\neq 0$), we have 
\begin{eqnarray}
\label{eq.DMA.34.0}
\overline{\langle n_{x}(t)\rangle - \langle n_{x}(\infty)\rangle}\approx
-\frac{\hat{\alpha}^{1/3}}{\overline{{\cal L}}\xi}\sqrt{\frac{8\pi^{3}\tau}{3}}
\exp\left[-\frac{3}{2}(4\pi^{2}\hat{\alpha}^{2}\tau)^{1/3}+\left(\frac{B+2|E|}{2|E|}\right)\xi^{2}\tau\right]
\end{eqnarray}

ii) In the other one, where $L^{2}\sim |E|t$, at $\hat{\beta}=1$, with $\hat{\alpha}\neq \xi\xi'$, according to (\ref{eq.1.14}), we have 
\begin{eqnarray}
\label{eq.DMA.35}
\overline{\langle n_{x}(t)\rangle - \langle n_{x}(\infty)\rangle}\approx
\frac{\hat{\alpha}}{\xi\overline{{\cal L}}}\left(\pi^{2}|E|t\right)^{\frac{1}{2}} \exp\left[-\frac{\hat{\alpha}}{\xi}(\pi^{2}|E|t)^{\frac{1}{2}}\right]
\left(\langle n_{x}(t)\rangle - \langle n_{x}(\infty)\rangle\right)
\end{eqnarray}

In this expression, $\langle n_{x}(t)\rangle - \langle n_{x}(\infty)\rangle$ 
is given in (\ref{eq.DMA.28}), for the case $|B|\neq C+D$.

Similar computations can be performed for  $\overline{\langle n_{x}(t)n_{x_{0}} \rangle -\langle n_{x}(\infty)n_{x_{0}} \rangle }$. For systems initially uncorrelated and homogeneous with initial density $\rho(0)$ and for the case $\hat{\beta}=1$, the long-time behaviour ($|E|t\gg 1, u=L^{2}/|E|t\ll 1$) of  the two-point correlation functions is the following: $\overline{\langle n_{x}(t)n_{x_{0}} \rangle -\langle n_{x}(\infty)n_{x_{0}} \rangle }\approx \tau^{\nu(\xi')}\exp\left(-\frac{3(\hat{\alpha}-\xi\xi')^{2/3}}{2}(4\pi^{2}\tau)^{1/3}+\left(\frac{B+2|E|}{2|E|}\right) \xi^{2} \tau\right)$, where $\xi'$ is the quantity defined above. We see that in absence of initial correlations, the  two-point correlation function has, in this regime, the same stretched-exponential long-time behaviour as the density (\ref{eq.DMA.34}). The exponent $\nu(\xi')$ depends on $\xi'$. For the cases considered here ($\hat{\beta}=1$) its value is $\nu=\frac{1}{6}$ (as for the SEP model with {\it free boundary conditions}, see below and Ref.\cite{Grynberg}) or $\nu=\frac{1}{2}$ .

In the other regime, the long-time behaviour ($|E|t\gg 1, u=L^{2}/|E|t\sim 1$) of  the two-point correlation functions 
 follows as in (\ref{eq.DMA.35}):$\overline{\langle n_{x}(t)n_{x_{0}} \rangle -\langle n_{x}(\infty)n_{x_{0}} \rangle }\approx \frac{\hat{\alpha}}{\xi\overline{{\cal L}}}\left(\pi^{2}|E|t\right)^{\frac{1}{2}} \exp\left[-\frac{\hat{\alpha}}{\xi}(\pi^{2}|E|t)^{\frac{1}{2}}\right]
\left(\langle n_{x}(t) n_{x_{0}}(0)\rangle - \langle n_{x}(\infty) n_{x_{0}}(0)\rangle\right)$  .

In the presence of initial correlations, the situation changes radically:  one should take into account the initial correlation in the disorder-averaging. The latter give rise to results which would be very different from the ones obtained for the density.

\section{Illustration: Two examples}

As a first example we consider the {\it symmetric exclusion model} (SEP) $A+\emptyset\leftrightarrow \emptyset +A$, with rates $\Gamma_{0 1}^{1 0}= \Gamma_{1 0}^{0 1}=\Gamma>0$, on a diluted chain (periodic and with free boundaries) with distribution probability: ${\cal P}(\Gamma)=q\delta_{\Gamma,\Gamma_{1}}+(1-q)\delta_{\Gamma,0}$. With the results of the section III.A, III.B, V and VI, setting $\hat{\alpha}=\hat{\beta}=1$, in the expression of ${\cal A}$ and ${\cal B}$ (see section V), in the regime $\Gamma t \gg L^{2}$, we recover the result of Grynberg and Stinchcombe\cite{Grynberg}: 
\begin{eqnarray}
\label{eq.0.50}
\overline{\langle n_{x}(t)n_{x_{0}}\rangle -\langle n_{x}(\infty)\rangle }\approx \rho(0)(1-\rho(0))\sqrt{\frac{2\pi}{3}}(4\pi^{2}\tau)^{1/6}\exp\left[-\frac{3}{2}(4\pi^{2}\tau)^{1/3}\right]
\end{eqnarray}
This result is valid for open segments as well as periodic segments.

 As second illustration, we consider the dynamics of the system described by the following processes: $A+A\leftrightarrow \emptyset+\emptyset$, with rates $\Gamma_{0 0}^{1 1}=\Gamma_{1 1}^{0 0}=\Sigma$ and $A+\emptyset\leftrightarrow \emptyset+A$, with rates $\Gamma_{1 0}^{0 1}=\Gamma_{0 1}^{1 0}=\Gamma$.

This model is relevant to the description of dimer adsorption and desorption (see Ref.\cite{Grynberg1}, and references therein).

Contrary to the SEP model, here the effect of open boundaries is real. We assume that $\Gamma_{1}>\Sigma_{1}$ and  that there is neither injection nor evacuation of particles at the boundary ({\it free boundary conditions}) of a segment (i.e. $\widetilde{\alpha}=\widetilde{\beta}=
\widetilde{\gamma}=\widetilde{\delta}=0 $).

This model preserves the {\it parity of the number of particles}, a property which  has to be taken into account when  computing the steady states (see footnote $2$).
 
Consider a diluted chain described by 
the distribution probability: ${\cal P}(\Gamma,\Sigma)=q\delta_{\Gamma,\Gamma_{1}}\delta_{\Sigma,\Sigma_{1}} + (1-q)\delta_{\Gamma,0}\delta_{\Sigma,0}\;\;, (0<q<1)$.

Thus, for segments with  {\it periodic boundary conditions}, the disorder-distribution simply reads $\omega({\cal L})=q^{{\cal L}}$.

If we assume  segments with {\it free-boundary conditions}, the disorder-distribution reads:

 $\omega({\cal L})=\left\{
\begin{array}{l l}
(1-q)^2 q^{{\cal L}}  &\mbox{, if $1\leq {\cal L} \leq L -2$ }\\
(1-q) q^{{\cal L}}   &\mbox{, if $ {\cal L} =L-1$  }
\end{array}
\right.$. 

It is useful to introduce the notations:
$\phi=\frac{1}{2}\;;\;0<\eta=\frac{\sqrt{\Gamma_{1}}-\sqrt{\Sigma_{1}}}{\sqrt{\Gamma_{1}}+\sqrt{\Sigma_{1}} }\leq 1\;;\; \triangle= 1+\frac{2\Sigma_{1}}{\Gamma_{1}-\Sigma_{1} }>0\nonumber\;;\;
B=-2(\Gamma_{1}+\Sigma_{1})\;;\;E=\Gamma_{1}-\Sigma_{1}\;;\;\xi'_{1}=0\;;\;$ and $\xi'_{2}=\ln \triangle \;;\;\xi'_{3}=\ln \eta\;;\;
\sigma_{1}=\sigma_{4}=\Sigma_{1}\;;\;\sigma_{2}=\sigma_{6}=\Gamma_{1}-\Sigma_{1}\;;\;\sigma_{3}=\sigma_{5}=\Gamma_{1}+\Sigma_{1}\;;\;\sigma_{7}=2\Gamma_{1}\;;\;\sigma_{8}=\sigma_{9}=2\Sigma_{1}$

We first consider the case of segments with periodic boundary conditions.
For a system with non-uniform initial conditions: $\langle n_{m}(0) \rangle=
\langle n_{m}(\infty)\rangle(1+\delta_{m,1})=\frac{1}{2}(1+\delta_{m,1})$, we obtain ($r=m-1$), in the regime where $\xi,(\Gamma_{1}-\Sigma_{1})t \gg 1$ , and $\tau\equiv\frac{(\Gamma_{1}-\Sigma_{1})t}{\xi^{2}}$.
\begin{eqnarray}
\label{eq.0.51}
\overline{\langle n_{r}(t)\rangle -\langle n_{r}(\infty)\rangle }\approx
 \langle n_{m}(\infty)\rangle\left(\frac{4\pi^{2}\tau}{\hat{\alpha}^{4}}\right)^{1/6}  \exp\left( -\frac{3}{2}((2\pi\hat{\alpha})^{2}\tau)^{1/3}-\frac{4\Sigma_{1}\xi^{2}\tau}{\Gamma_{1}-\Sigma_{1}}\right)
\end{eqnarray}

Let us now pass to an open  system, with free boundary conditions, for which the initial state is uncorrelated with density
 $\langle n_{m}(0) \rangle=\rho(0)$, (with $\rho(0)\neq 0$), we have:

In the regime where $|\Gamma_{1}|t,|\Sigma_{1}|t\gg L^{2}\gg 1$, with $\xi'=\ln \triangle >0$, ($\hat{\alpha}\neq\xi\xi'$)

\begin{eqnarray}
\label{eq.0.51}
\overline{\langle n_{x}(t)\rangle - \langle n_{x}(\infty)\rangle}\approx
\left\{
\begin{array}{l l}
\left(\frac{4\pi^{2}\tau}{(\hat{\alpha}-\xi\ln \triangle)^{4}}\right)^{1/6}\exp\left(-\frac{3(\hat{\alpha}-\xi\ln \triangle)^{2/3}}{2}(4\pi^{2}\tau)^{1/3}-\frac{4\Sigma_{1}\xi^{2}\tau}{\Gamma_{1}-\Sigma_{1}}\right)  &\mbox{, if $\triangle <e^{2\hat{\alpha}/\xi}$ }\\
-\hat{\alpha}^{1/3}\sqrt{\frac{8\pi^{3}\tau}{3}}
\exp\left[-\frac{3}{2}((2\pi\hat{\alpha})^{2}\tau)^{1/3}
-\frac{4\Sigma_{1}\xi^{2}\tau}{\Gamma_{1}-\Sigma_{1}}\right]  &\mbox{, if $\triangle \geq e^{2\hat{\alpha}/\xi} $  }
\end{array}
\right.
\end{eqnarray}
We note  on the basis of this simple example, the effects of the boundary conditions. 
The  SEP model, where the long-time dynamics does not depend on the boundary
 conditions (see Ref.\cite{Grynberg}) of the segments is a special case.
\section{Summary and conclusion}

We have considered the single-species one-dimensional models  characterized by rates lying on a $10-$parametric manifold. Such reaction-diffusion processes are (in arbitrary dimensions) formally completely soluble because the equations of motion of correlation functions are closed. 
After having generalized a previous approach to the  dynamics of (ordered) systems with {\it open boundaries}, with help of the {\it Dynamical Matrix Ansatz}, we investigated
the dynamics of this class of processes in the presence of {\it quenched} disorder (consisting in a random distribution of the reaction of {\it periodic} and {\it open} segments). We have computed the density  and the two-point non-instantaneous correlation functions for various initial states. The study of the dynamics in the presence of quenched disorder leads to various regimes. The most remarkable effect of the disorder appears when time is much larger than the typical diffusion time. In this case the finite-size effect radically changes the dynamics: for the SEP model, the non-instantaneous correlation functions no longer decay algebraically but follow rather a stretched exponential. For the other models, in this regime, the {\it quenched disorder} also affects the dynamics but less {\it dramatically}: the decay of the density and two-point correlation functions is a combination of the {\it ordered} exponential (corrected by subdominant terms arising from the disorder average) and a power-law arising from the {\it quenched disorder} average.
Similar effects of disorder have been reported in recent experiments \cite{Kuroda}. We also pointed out the {\it non-trivial} effect of the boundary conditions on the disorder average.
We have illustrated our results by considering two physically relevant examples.
\\

The support of Swiss National Fonds is gratefully acknowledged.
%
%

%
%
\end{document}